# Origin of Large Dielectric Constant with Large Remnant Polarization and Evidence of Magnetoelectric Coupling in Multiferroic La modified $BiFeO_3$-$PbTiO_3$ Solid Solution

Anupinder Singh, Ratnamala Chatterjee*

Magnetics & Advanced Ceramics Laboratory, Physics Department, Indian Institute of Technology Delhi, Hauz Khas, New-Delhi-110016

S. K. Mishra, P.S.R. Krishna and S.L. Chaplot

Solid State Physics Division, Bhabha Atomic Research Centre, Trombay, Mumbai-400085, India

**ABSTRACT:**

The presence of superlattice reflections and detailed analyses of the powder neutron and x-ray diffraction data reveals that La rich $(BF_{0.50}–LF_{0.50})_{0.50}–(PT)_{0.50}$ (BF-LF-PT) has ferroelectric rhombohedral crystal structure with space group *R3c* at ambient condition. The temperature dependence of lattice parameters, tilt angle, calculated polarization ($P_s$), volume, and integrated intensity of superlattice and magnetic reflection show an anomaly around 170 K. Extrapolation of $P_s$ vs. $T$ curve indicates that the ferroelectric transition temperature in this La rich BF-LF-PT solid solution is at $T_c^{FE}$ = 523 K. Impedance spectroscopy, dielectric spectroscopy and ac conductivity measurements were performed in temperature range $473K \leq T \leq 573K$ to probe the origin of large remnant polarization and frequency dependent broad transitions with large dielectric constant near $T_c^{FE}$. Results of impedance spectroscopy measurements clearly show contributions of both grain and grain boundaries throughout the frequency range ($10^2$ Hz $\leq f \leq 10^7$ Hz). It could be concluded that the grain boundaries are more resistive and capacitive as compared to the grains, resulting in inhomogeneity in the sample causing broad frequency dependent dielectric anomalies. Enhancement in dielectric constant and remnant polarization values are possibly due to space charge polarization caused by piling of charges at the interface of grain and grain boundaries.The imaginary parts of dielectric constant ($\varepsilon''$) vs. frequency data were fitted using Maxwell-Wagner model at $T_c^{FE}$ (~523 K) and model fits very well up to ~$10^5$ Hz. Magnetodielectric measurements prove that the sample starts exhibiting magnetoelectric coupling at ~170 K, which is also validated by neutron diffraction data.

***Corresponding Author: ratnamalac@gmail.com, rmala@physics.iitd.ac.in**

## I. INTRODUCTION

Multiferroics represent an appealing class of functional materials that exhibit different ferroic interactions simultaneously. The coexistence of several interactions, particularly existence of a cross-coupling between the magnetic and electric interactions, termed as magnetoelectric (ME) coupling, brings about novel physical phenomena and offer possibilities for new device functions [1-3]. $BiFeO_3$ (BF) is one of the most extensively studied multiferroic material in recent years [2-6] and it is the only material known to exhibit magnetic order ($T_N$= 643 K) and ferroelectric order (ferroelectric transition temperature $T_c^{FE}$= 1103 K) at room temperature. At 300 K, It has a rhombohedral structure with space group *R*3*c*, which permits coupling between magnetism and ferroelectricity. It shows G-Type antiferromagnetic spin configuration along $[001]_h$ direction with a long range (periodicity ~ 620 Å) cycloidal spin structure incommensurate with the lattice along $[110]_h$ direction of the hexagonal unit cell of the rhombohedral structure below magnetic ordering temperature. It also exhibits a very weak magnetoelectric coupling [7,8] due to cycloidal spin modulation which thought to be the reason for suppressed linear magnetoelectric coupling. Theoretical calculations [9] predict that a large difference between the transition temperatures $T_c^{FE}$ and $T_N$ causes weak ME coupling in BF. In addition to this, BF also has other shortcomings like its semiconducting behavior at $T \geq 300$ K, which does not allow electric poling and causes high dielectric losses in the sample at room temperature. Due to this, it is difficult to measure the ferroelectric properties of BF at and above room temperatures and limits its applications in devices. Thus the main aim of researchers in the field has been **(i)** to improve the dielectric properties of BF/ modified BF and **(ii)** to bring ferroelectric transition ($T_c^{FE}$) and magnetic transition ($T_N$) temperatures in the modified BF compositions closer to each other; thus achieving an enhancement of ME coupling.

The incommensurate cycloidal spin structure can be suppressed with application of (a) magnetic field, (b) strain and probably (c) chemical substitution too [1-3]. In 1996, Sosnowska *et al.* [10] demonstrated that the La doping at A-site in $BiFeO_3$ destroys the space modulated spin structure (SMSS) and allows the measurement of linear ME effect. In 2006, Zhang *et al* [11] reported enhancement of magnetic and ferroelectric properties for La substituted BF. In addition to this, various workers have shown that La substitution decreases the ferroelectric transition temperature in materials belonging to perovskite family **[12]**. In 2005, Wang *et al.* [13] reported an enhancement of magnetoelectric properties in the (1-$x$) $BiFeO_3$-($x$) $PbTiO_3$ (BF-PT) solid solutions and also claimed destruction of space modulated spin structure in them, due to addition of $PbTiO_3$. Using detailed powder x-ray diffraction studies, Zhu *et al.* [14] proposed phase diagram for the solid solution of (1-$x$) $BiFeO_3$-($x$) $PbTiO_3$(BF-PT). Their results reveal the existence of a morphotropic phase boundary (MPB) in this system, in which rhombohedral ($x \leq 0.20$), orthorhombic ($0.20 \leq x \leq 0.28$) and tetragonal ($x \geq 0.31$) phases exist with a large tetragonality in the tetragonal phase region. In the MPB region, they reported the simultaneous existence of orthorhombic, tetragonal and rhombohedral phases. It is important to note that they ruled out the presence of monoclinic phase at morphotropic phase boundary region. They also pointed out that in BF-PT solid solutions, compositions with BF:PT ratio around 50:50 favors the formation of chemically ordered microregions in which spiral spin modulation of BF is not only disturbed but completely disappears and may give rise to weak ferromagnetic state. Further, Cheng *et al.* [15] reported that tetragonality of BF-PT solid solution decreases as concentration of the $La^{3+}$ ions increases (for $x< 0.3$). They suggest (but without any experimental evidence), that further increase of La content in BF-PT matrix may stabilize rhombohedral (*R3c*) phase. They also reported improved magnetic properties in $(BF_{1-x}–LF_x)–(PT)$ system at MPB (45:55). Recently, detailed studies on the effect of further increasing La content in $(BF_{0.50}–LF_{0.50})_{0.50}–(PT)_{0.50}$ (BF-LF-PT) has been investigated by the authors [16]. Our studies on La rich BF-LF-PT solid solution revealed that large grains (~7μm) could be obtained by optimizing the processing parameters, and the sample properties were significantly improved. The sample showed improved spontaneous polarization, magnetization and magnetoelectric coupling. The dopants help in bringing the ferroelectric and magnetic transition temperatures closer to each other; a condition favored for technological applications of the materials.

However, the complex mechanism of origin of spontaneous polarization ($P_s$) and magnetization in La rich BF-LF-PT still needs to be understood and requires further probing. Some open questions that require further explanations are: **(a)** the observation of large dielectric constant and ferroelectric hysteresis *(P-E)* loop in the sample is contradictory to the observation of centrosymmetric cubic crystal structure with space group *Pm3m*, **(b)** the true polarization behavior of the sample needs to be understood to explain **(i)** the observed frequency dependent broad transitions with large dielectric constant near $T_c^{FE}$ **(ii)** large remnant polarization observed in *P-E* loops, and **(c)** a non saturating magnetic *(M-H)* hysteresis loop observed at room temperature could not establish the existence of long-range magnetic order in this La rich BF-LF-PT system[16] .

In this paper, we present the results of powder diffraction (temperature dependent neutron diffraction, rotating anode powder x-ray diffraction at room temperature) in conjunction with detailed impedance spectroscopy (dielectric spectroscopy, ac conductivity measurements). The results observed from these measurements would be used to clarify first two issues ((a) and (b)) addressed above. The goal of impedance spectroscopy investigation is to improve our understanding of these La rich BF-LF-PT properties by analyzing the electrical response of the grain and grain boundary effects, making an interpretation of the microscopic process. However, the powder x-ray and neutron diffraction study helps to assign correct crystal structure of the system. Neutron diffraction offers certain unique advantages over x-rays, especially in the accurate determination of light elements like oxygen positions. It provides information about subtle changes in structure associated with the oxygen atoms and accompanying phase transitions that are crucial for resolving the above mentioned existing controversies/ issues. The observation of an anomaly at 170 K, in the temperature dependence of lattice parameters, magnetization and polarization vs. temperature is taken as evidence of multiferroics magnetoelectric coupling. Magnetodielectric measurements (10 K$\leq T \leq$300 K) finally prove that the sample starts exhibiting magnetoelectric coupling in this temperature range at ~170 K.

## II. EXPERIMENT

### (a) Sample preparation and characterization

Solid solutions are prepared using a conventional solid state reaction route as described elsewhere [16]. Powder X-ray diffraction studies were carried out using 12kW rotating Cu anode based Rigaku powder diffractometer operating in the Bragg-Brentano focusing geometry. Data were collected in the continuous scan mode at a scan speed of 1 degree per minute and step interval of 0.02 degree. The powder neutron diffraction data were recorded in the 2θ range of 7°–138° at a step width of 0.05° using neutrons of wavelength of 1.249 Å on a medium resolution powder diffractometer in the Dhruva Reactor at Bhabha Atomic Research Centre. The structural refinements were performed using the Rietveld refinement program FULLPROF **[17]**. A Thompson-Cox-Hastings pseudo-Voigt with Axial divergence asymmetry function was used to model the peak profiles. The background was fitted using a sixth order polynomial. Except for the occupancy parameters of the atoms, which were fixed corresponding to the nominal composition, all other parameters, i.e., scale factor, zero displacement, isotropic profile parameters, lattice parameters, isotropic thermal parameters and positional coordinates, were refined.

The impedance ($Z$), phase angle ($\varphi$), capacitance, dielectric loss ($\tan\delta$), and *ac* conductivity were measured in the frequency range $10^2 \leq f \leq 10^7$ Hz using an HP4192A impedance analyzer. The temperature dependent impedance measurements were carried out at an interval of 10 K in the range 473 K to 573 K and the temperature was controlled with an accuracy of ±1K using a temperature controller. The low temperature (5 K $\leq T \leq$ 250 K) dc and ac magnetic measurements were carried out using SQUID magnetometer (MPMS-XL7). While, low temperature dielectric constant ($\varepsilon'$) and magneto-dielectric constant vs. temperature measurements were carried out using LCR meter and a cryogen free low temperature high magnetic field facility.

### (b) Data Analysis

The real (Z') and imaginary (Z") parts of the complex impedance ($Z^*$) were calculated from the measured data using following relations.

$$Z' = Z * \cos(\varphi) \tag{1}$$

$$Z'' = Z * \sin(\varphi) \tag{2}$$

Where complex impedance is given by the relation

$$Z^* = Z' + iZ'' \tag{3}$$

The real ($\varepsilon'$) and imaginary ($\varepsilon''$) parts of dielectric constant are calculated using following relations:

$$\varepsilon' = \frac{C.t}{\varepsilon_0 A} \tag{4}$$

$$\varepsilon'' = \varepsilon'(\tan\delta) \tag{5}$$

Where $C$ = measured capacitance, $t$= thickness of sample, $\varepsilon_0$ = permittivity of free space, $A$ = cross sectional area of electrode surface and tan $\delta$ = dielectric loss.

## III. RESULTS

### A. *Structural analysis: A powder x-ray and neutron diffraction studies*

Figure 1 (a) depicts the Rietveld refinements of powder x-ray diffraction patterns of BF-LF-PT system at room temperature using cubic symmetry and with space group *Pm3m* and all the peaks can be indexed. But, the presence of ferroelectric hysteresis loop [16] contradicts the observation and suggests non-centrosymmetric nature of crystal structure at ambient conditions. The intensity of the superlattice reflection resulting from octahedral tilts depends on the small changes in the position of oxygen atoms only. The x-ray scattering form factor for oxygen is rather small, and if the tilt angles are small, the superlattice reflections will have too low an intensity to be visible above the background count level. Thus, the non-observation of superlattice reflections in a powder x-ray diffraction pattern may lead to misinterpretation. In view of this, we re-visited the crystal structure for the BF-LF-PT system using powder neutron diffraction technique.

To check the correctness of structure, first, we refined powder neutron data using cubic symmetry (space group *Pm3m*) as shown in Figure 1(b). It is evident from the figure that all the peaks of powder neutron diffraction data cannot be indexed. The origin of these additional Braggs peaks may be either nuclear (structure) or magnetic (in case of antiferromagnetic). The presence of additional Braggs peaks with well defined intensity at high angles clearly ruled out the possibility of only magnetic origin and clearly suggests that the cubic structure is incorrect. It is important to bring to notice that presence of superlattice reflection around ~ 38 degree is observed even in powder x-ray diffraction data [when data was collected in step scan mode and shown in inset of figure 1(a)]. These observations confirm the nuclear origin of additional Braggs peaks and non-centrosymmetric structure for BF-LF-PT. In view of the observation of ferroelectric hysteresis loop at room temperature, we also refine powder neutron diffraction data using different classical perovskite space groups that exhibit ferroelectric properties, like tetragonal *P4mm*, orthorhombic *B2mm*, rhombohedral *R3m* and monoclinic *Cm* or *Pm*; none of these space groups account the additional Braggs peaks (also termed as superlattice reflections).

The Miller indices of the superlattice reflections based on a doubled pseudo-cubic cell carry information about the nature of the octahedral tilts in the structure. Superlattice reflections with all-odd integered indices ("odd-odd-odd" i.e., "ooo" type) and two-odd and one-even integered indices result from anti-phase (− tilt) and in-phase (+ tilt) tilting of the adjacent oxygen octahedra due to structural phase transitions driven by softening and freezing of the phonons at *R* (*q*= ½ ½ ½ ) and *M* (q= ½ ½ 0) points of the cubic Brillouin zone, respectively [18]. The ferroelectric instabilities in perovskite are associated with Γ point. The superlattice reflection observed by us in powder x-ray diffraction has odd-odd-odd type (i.e. 311) integred indices, when indexed with respect to doubled pseudo-cubic cell and confirms the *R* point instabilities. The onset of ferroelectric polarization transforms under $Pm\bar{3}m$ as the components of the $\Gamma_4^-$ irreducible representation (IR). Anti-phase tilting of oxygen octahedra transforms under $Pm\bar{3}m$ as the components of the $R_4^+$ IR. Using the software package "ISOTROPY 2000 [19]" to couple $\Gamma_4^-$ and $R_4^+$ IRs, we obtained many space groups. We find that the rhombohedral structure with space group *R3c* accounted all the reflection at room temperature. In the space group *R3c*, for the description of the crystal structure, we used the following hexagonal axes: $\mathbf{a_h}=\mathbf{b_p}-\mathbf{c_p}$, $\mathbf{b_h}=\mathbf{c_p}-\mathbf{a_p}$ and $\mathbf{c_h}=2\mathbf{a_p}+2\mathbf{b_p}+2\mathbf{c_p}$. The hexagonal unit cell contains 6 formula units, while the true unit cell in the primitive rhombohedral contains 2 formula units. The asymmetric unit of the structure consists of Bi/Pb/La atom at the 6*a* site at (0,0,1/4+*w)*, Fe/Ti atom at the 6*a* site at (0,0,0+*w*), and O atom at the 18*b* site at (1/6 +*u*,1/3+*v*,1/12). The fit between the observed and calculated profiles is quite satisfactory and include the weak superlattice reflections (see Fig 1(c)).

Figure 2 depicts the evolution of powder neutron diffraction patterns of the sample with temperature in the range of 25–300 K for the two-theta range 10 to 80 degree. It is evident from this figure, in addition to superlattice peak, one additional peak around 2θ= 15.58 degree (similar to BF) appears below 200 K and its intensity increases with deceasing temperature and suggests that magnetic origin. Variation of the integrated intensity for the one of strong magnetic reflection (at $2\theta \approx 15.58$ degree) and superlattice reflection (at $2\theta \approx 30.25$) as a function of temperature are shown in figure 3 and show change of slope around 200 K. The other interesting feature noted after Rietveld refinements is that the calculated equivalent lattice parameters (Fig.4(a)), volume (Fig.4(b)), tilt angle (Fig.4(c)), and calculated polarization (Fig.4(d)) as a function of temperature, indicate a change between 150 K-200 K, suggesting some structural anomaly in this temperature range.

In the next sections we discuss the results of impedance spectroscopy and ac conductivity behavior of the sample, to understand the origin of frequency dependent broad transitions observed in this sample and large remnant polarization values.

### B. ***Relaxation Mechanism***

#### ***(1) Impedance Spectroscopy***

Impedance spectroscopy is a well known tool for detecting the dielectric relaxation processes in a polycrystalline sample. It involves the application of an alternating voltage signal to a sample and measurement of phase shifted current response. It is established that the complex impedance spectroscopy is a powerful tool to analyze the relationship between microstructure and properties. Using impedance spectroscopy one can distinguish between intrinsic (grain), extrinsic (grain boundaries) and sample electrode interface contributions to observed properties of the sample. An equivalent circuit based on impedance spectra gives an insight into the physical process occurring inside the sample.

Figures 5(a) and 5 (b) show the variations of the real ($Z'$) and imaginary parts ($Z''$) of impedance as a function of frequency at different temperatures (473K≤*T*≤573K). Two clear broad peaks are observed in both Z′ and Z″ profiles and the intensity of these peaks in both profiles are seen to decrease with increasing temperature. As is evident in Figs. 5(a) and 5(b) both peaks shift towards higher frequencies with rise in temperature; first peak at ~$10^3$ Hz for 473 K data shifts to ~$10^5$ Hz at 573 K whereas, the second peak observed at ~$10^6$ Hz for 473 K data is seen to shift to ~$10^7$ Hz at 573 K.

The activation energy associated with both relaxation processes are determined by fitting (see Fig.5(c)) the Arrhenius equation in temperature range (473K ≤T≤ 573K) using the relation

$$\tau = \tau_0 \exp\left(\frac{-E_a}{k_B T}\right), \qquad \tau = \frac{1}{2\pi f_p} \tag{6}$$

where $\tau_0$ is the prefactor, $E_a$ is the activation energy for relaxation process and $k_B$ is Boltzamnn constant. The fitting parameters thus obtained are $\tau_0$ (grain) = $4.2\times10^{-13}$s, $E_a$ (grain)=0.596eV, $\tau_0$ (grain boundary)= $4.21\times10^{-15}$s and $E_a$(grain boundary)=1.013eV.

Figure 5(d) shows Nyquist plots ($Z''$vs.$Z'$) at all temperatures (473K≤$T$≤573K) having two semi circle arcs (see inset of 5(d)). In Nyquist plots, these semi circle arcs represent different relaxation processes. Usually semi circle at lower frequency represents grain boundary relaxation process and the one at higher frequencies gives the information on grain relaxation. It is evident from Nyquist plots that both grain and grain boundaries are contributing to the physical properties of this polycrystalline sample. However, it is observed that the centers of these semicircles arcs are below the real impedance axes. The complex impedance can be represented by Cole- Cole equation.

$$Z^* = \frac{R_g}{1+(j\omega R_g C_g)^{\alpha_g}} + \frac{R_{gb}}{1+(j\omega R_{gb} C_{gb})^{\alpha_{gb}}} \tag{7}$$

Where $Z^*$ = Complex Impedance, $R_g$= Grain Resistance, $R_{gb}$= Grain Boundary Resistance, $C_g$= Grain Resistance, $C_{gb}$= Grain Boundary Resistance, $\sigma_g$ = relaxation time distributed function for grain, $\sigma_{gb}$ = relaxation time distributed function for grain boundary. "*j*" and $\omega$ have their usual meanings. In this equation the relaxation time distributed function 'α' is given by the relation "$\theta = \frac{\alpha\pi}{2}$". The value of 'α' varies from zero to unity; for pure Debye process the value for 'α' should be unity. Thus 'α' is a measure of deviation of relaxation process from the ideal Debye behavior. The values of depression angle ($\theta$) (angle between the real impedance axes and the centre of semicircle arc) vary in the range (~9°-19°) and calculated '$\alpha$' values (~0.10-0.35) indicate that both grain and grain boundary relaxation processes are non Debye type. In order to account for non-Debye behavior, a well established approach [20] of replacing the specific capacitance '*C*' by a phenomenological constant phase element (*CPE*) was used. The impedance of CPE is given by $Z_{CPE} = \frac{1}{(j\omega)^{\beta} CPE}$, where β < 1. The Nyquist plots are well fitted with *R-CPE* model (shown in Fig.6 (a)) both for grain and grain boundaries. A representative plot and equivalent circuit fitting are shown in Fig 6(b). The fitting parameters R, $C$ and $\beta$ both for grain ($R_g$, $C_g$, $\beta_g$) and grain boundaries ($R_{gb}$, $C_{gb}$, $\beta_{gb}$) are obtained and variation of these parameters with temperature is shown in Figs 6(e), (f) and (g). The grain boundaries in this sample are found to be more capacitive and resistive than the grains. Activation energies determined by fitting these resistance values (grain and grain boundaries) in Arrhenius equation (Fig 6(e)) are in agreement with the $E_a$ values obtained from relaxation time Arrhenius fit. The values of parameter $\beta$ (see Fig.6 (g)) once again confirm that both relaxation processes are non Debye type. The lower activation energy for grains confirms that the grains are more conducting than the grain boundaries, as is clearly evident from the equivalent circuit fitting.

***(2) Dielectric Spectroscopy***

The variation of real ($\varepsilon'$) and imaginary ($\varepsilon''$) parts of dielectric constant as a function of frequency ($10^2 Hz \leq f \leq 10^7 Hz$) at different temperatures ($473K \leq T \leq 573K$) are shown in Figs. 7(a) and (b). At lower frequencies both $\varepsilon'$ and $\varepsilon''$ have large values that decrease with increase in frequency. In dielectric materials the coexistence of low dielectric constant regions with high dielectric constant region can influence the piling of charges at the interface and result in Maxwell-Wagner (M-W) polarization [21]. Catalan *et al.* [22,23] reported the onset of Maxwell-Wagner polarization at lower frequencies in ferroelectric super lattices. The impedance spectroscopy results clearly validate the existence of less resistive and capacitive grains and highly resistive and capacitive grain boundaries in the sample that may result in dielectric inhomogeneity, resulting in M-W polarization in our sample. We fit Maxwell-Wagner model to $\varepsilon''$ vs. frequency profile (see Fig. 7(c)) using following relations [23].

$$\varepsilon''(\omega) = \frac{1}{\omega C_0 \left(R_{gb} + R_g\right)} \frac{1 - \omega^2 \tau_{gb} \tau_g + \omega^2 \tau (\tau_{gb} + \tau_g)}{1 + \omega^2 \tau^2} \tag{8}$$

where subindex '$g$' and '$gb$' refer to grain and grain boundary respectively.

$$\tau_g = C_g R_g \,,\ \tau_{gb} = R_{gb} C_{gb} \,,\ \tau = \frac{\tau_{gb} R_g + \tau_g R_{gb}}{R_g + R_{gb}} \,,\ C_0 = \varepsilon_0 \frac{A}{t} \tag{9}$$

Up to ~$10^5$ Hz frequencies the Maxwell-Wagner model fits very well with observed data implying that at lower frequencies the large dielectric values are due to Maxwell-Wagner interfacial polarization. At higher frequencies, where space charge polarization is known to be absent $\varepsilon''$ values show deviation from the model.

***3. Electric conduction Mechanism***

The *ac* conductivity response with frequency was also studied in the sample, to confirm the Maxwell-Wagner polarization at lower frequencies (<$10^5$ Hz). Frequency dependent ac conductivity ($\sigma_{ac}$) was measured in the range $10^2 Hz \leq f \leq 10^5 Hz$ at temperatures ranging from *473K* to *573K* and the results are shown in the Figs. 8 (a) and (b). Temperature range is typically chosen in the vicinity of dielectric anomaly. A plateau, or a frequency independent behavior of *ac* conductivity at lower frequencies indicate the dominance of *dc* conductivity to conduction mechanism and the absence of hopping charge carrier polarization in this frequency range. The frequency dependence is analyzed by using "Jonscher's Universal Power Law" given by the relation [24]

$$\sigma_{ac} = \sigma_{dc} + A\omega^n$$

where $\sigma_{dc}$ is *dc* conductivity at particular temperature, *A* is a temperature dependent constant and *n* is the power law exponent which generally varies between 0 and 1 depending upon temperature. The exponent *n* represents the degree of interaction between mobile ions with lattice around them (*n*=1 represents non-interacting Debye system and with decreasing *n*, interaction between mobile ions and lattice is expected to increase). Pre-factor *A* gives the strength of polarizibility. The universal power law fits well up to ~$10^5$ Hz frequency at all temperatures (see Figs. 8(a) and (b)) and fitting parameters $\sigma_0$, *A*, and *n* are thus obtained.

The variations of the pre factor *A* and exponent *n* as a function of temperature are shown in Figs. 8(c) and 8(d). Pre-factor *A* shows a peak around ~500 K, beyond which the values of '*A*' decreases drastically. The exponent *n* shows a minimum exactly around this temperature. There are anomalies in '*A*' and '*n*' are thus observed in the vicinity of dielectric anomaly.

**IV DISCUSSION**

The presence of superlattice reflections in powder x-ray and neutron diffraction patterns unambiguously show that the crystal structure is not centrosymmetric cubic (with space group *Pm*3*m*) but non-centrosymmetric rhombohedral (space group *R3c*) at room temperature, which is consistent with the observed ferroelectricity [16] in the sample. In addition to this, we also observed the anomalies in the lattice parameters, tilt angle, calculated polarization, volume, and integrated intensity of superlattice and magnetic reflection as a function of temperature around 170K.

The spontaneous polarization value (~16 μC/cm$^2$) as calculated from unit cell refinement, was found to be much lower than the values of polarization (~61μC/cm$^2$) earlier reported by the authors [16] at room temperature. However, the extrapolation of spontaneous polarization ($P_s$) vs. temperature curve (Fig.3(d)) shows that $P_s$ vanishes at ~523K, which is also the peak temperature ($T_c^{FE}$ ) in dielectric response of the sample. The polarizibility factor *A* obtained as a fitting parameter of frequency dependent ac conductivity curves, also shows a maximum around this temperature. All these factors clearly confirm a ferroelectric phase transition at around this temperature. The cause of such enhanced dielectric constant values at $T_c^{FE}$ and large remnant polarization values, as observed in the sample, (and also usually reported [21-23] in this field of research), are analytically argued using the detailed data analyses of dielectric measurements, impedance spectroscopy and *ac* conductivity measurements in the temperature range 473 K≤ T ≤573 K. Impedance spectroscopy in the range of 473K≤*T*≤573K, gives a clear picture of presence of two types (grains and grain boundaries) of relaxation processes in this sample. The suppressed semicircle arcs in Nyquist plots indicate that the relaxation processes for both grain and grain boundaries are non-Debye type. This means that there is heterogeneity in the sample and sample has a distribution of relaxation times around a mean relaxation time $\tau_m$. The equivalent circuit fitting for Nyquist plots with *R-CPE* model also confirms the non Debye relaxation processes present in the sample. The fitting parameters $R_g$, $R_{gb}$, $C_g$ and $C_{gb}$ as obtained from Nyquist plots show that in this sample the grain boundaries are more capacitive and resistive as compared to the grains. The estimated activation energy ($E_a$) obtained by Arrhenius fittings of mean relaxation times and resistances, indicate that the conduction process in grains is easier than that in grain boundaries.

The sample can be modeled as consisting of two regions (grains and grain boundaries) of different dielectric constants and conductivities, represented closely by Maxwell-Wagner model. The $\varepsilon''$ vs. frequency profiles fit very well up to ~$10^5$ Hz, verifying Maxwell-Wagner polarization in the sample (See Fig. 7(c)). The conducting grains and highly insulated grain boundaries result in piling of charge at interfaces, giving rise to large dielectric and polarization values in the frequency range up to ~$10^5$ Hz. The parameters used for the fitting of Maxwell-Wagner model are in agreement with the equivalent circuit fitting parameters of Nyquist plots. Obviously, the origin of the frequency dependent broad transitions (relaxor-like) in temperature range $473\text{K} \leq T \leq 573\text{K}$, and large dielectric constant values and large remnant polarization are argued to be the artifact of Maxwell Wagner polarization.

We also show that the dc conductivity dominates the conduction mechanism in the sample at lower frequencies and conduction due to hopping charge carriers is ruled out in this frequency range, giving full support to the Maxwell-Wagner polarization up to ~$10^5$ Hz.

Another important offshoot of this work is the evidence of M-E coupling in this La rich BF-LF-PT sample starting from T≥170 K. As shown in Fig. 3 all the calculated parameters of the sample including the calculated polarization from the temperature dependent neutron diffraction data showed a distinct change in the temperature range 150 – 200 K, suggesting the existence of some kind of structural anomaly in this sample around this temperature. Low temperature dielectric constant ($\varepsilon'$) vs. temperature (10 K ≤ T ≤ 300 K) profile was measured and analyzed specifically to identify any change in dielectric constant ($\varepsilon'$) in this temperature range (~150 K-200 K). A slope change in dielectric constant ($\varepsilon'$) vs. temperature profile in this temperature range is clearly evident in Fig.9(a). Interestingly both *dc* magnetization (*M*) and *ac* susceptibility measurements ($\chi''$) vs. temperature profiles preformed at 3Hz and 999Hz, in the low-temperature region show a slope change around same temperature (Figs.9(b) & (c)). Such slope changes in magnetization in BF-PT samples at low temperatures have been identified in literature [14] as Neel temperature $T_N$. However, the neutron diffraction data reported in this work clearly links these slope changes with a structural anomaly present in this $BiFeO_3$ (BF) based solid solution.

The results of Magnetodielectric measurements in the temperature range 10 K ≤ $T$ ≤ 300 K at $10^5$ Hz are shown in Fig. 10 Such high frequency is used to avoid the artifacts of Maxwell-Wagner Polarization. It is observed that although below 170 K the data in zero field and in field (3T) overlap completely, above 170 K an enhancement of dielectric constant ($\varepsilon'$) on application of magnetic field continues up to 300 K. Room temperature multiferroic behavior in this sample has already been reported by authors [16]. We thus conclude that M-E coupling in this La rich BF-LF-PT sample is observed in the temperature range 170 K ≤ $T \leq 300$ K.

## V. CONCLUSION

In summary, the presence of superlattice reflection in powder diffraction patterns confirm the noncubic structure of La modify BF-PT. Detailed Rietveld analysis of the neutron diffraction data reveals that it has ferroelectric rhombohedral crystal structure with space group *R3c* at the room temperature. The contradiction observed between the calculated spontaneous polarization (from temperature dependent neutron diffraction data) and measured remnant polarization values with the enhancement of dielectric constant values at $T_c$ can be understood on the basis of Maxwell–Wagner model. The sample can be modeled as consisting of two regions (grains and grain boundaries) of different dielectric constants and conductivities, represented closely by Maxwell-Wagner model. Both grain and grain boundaries show non Debye relaxation behavior. A substantial evidence of magnetoelectric coupling in the sample is established in the temperature range (170 K ≤ $T$ ≤ 300 K).

## ACKNOWLEDGEMENTS:

The authors would like to acknowledge SQUID National Facility at IIT Delhi (Funded by Department of Science and Technology (DST), India) for magnetic measurements. One of the authors (A.S) would like to acknowledge University Grants Commission (UGC), India for providing the fellowship. We would also like to acknowledge Dr. S. Patinaik, JNU, India for magnetodielectric measurements.

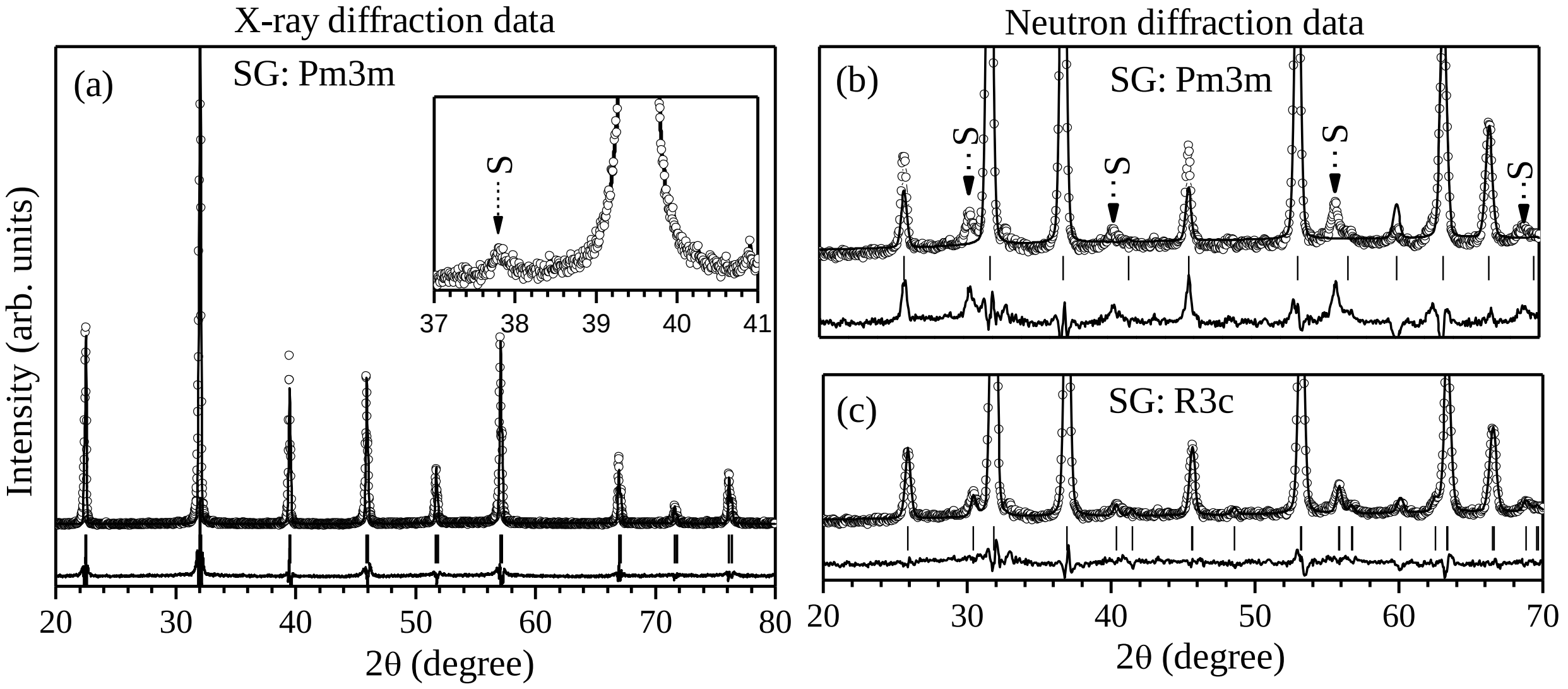

**Fig.** 1 Observed (dot), calculated (continuous line), and difference (bottom line) profiles obtained after the Rietveld refinement of BF-LF-PT with cubic space group ***Pm3m*** using (a) x-ray diffraction data (b) neutron diffraction data and (c) rhombohedral space group ***R3c*** using neutron diffraction data. The superlattice reflection (S) associated with tilting of octahedra is marked with arrow in powder diffraction patterns. Inset of figure (a) shows the presence of superlattice reflection (S) in x-ray diffraction pattern. Tick marks above the difference profile show peak positions.

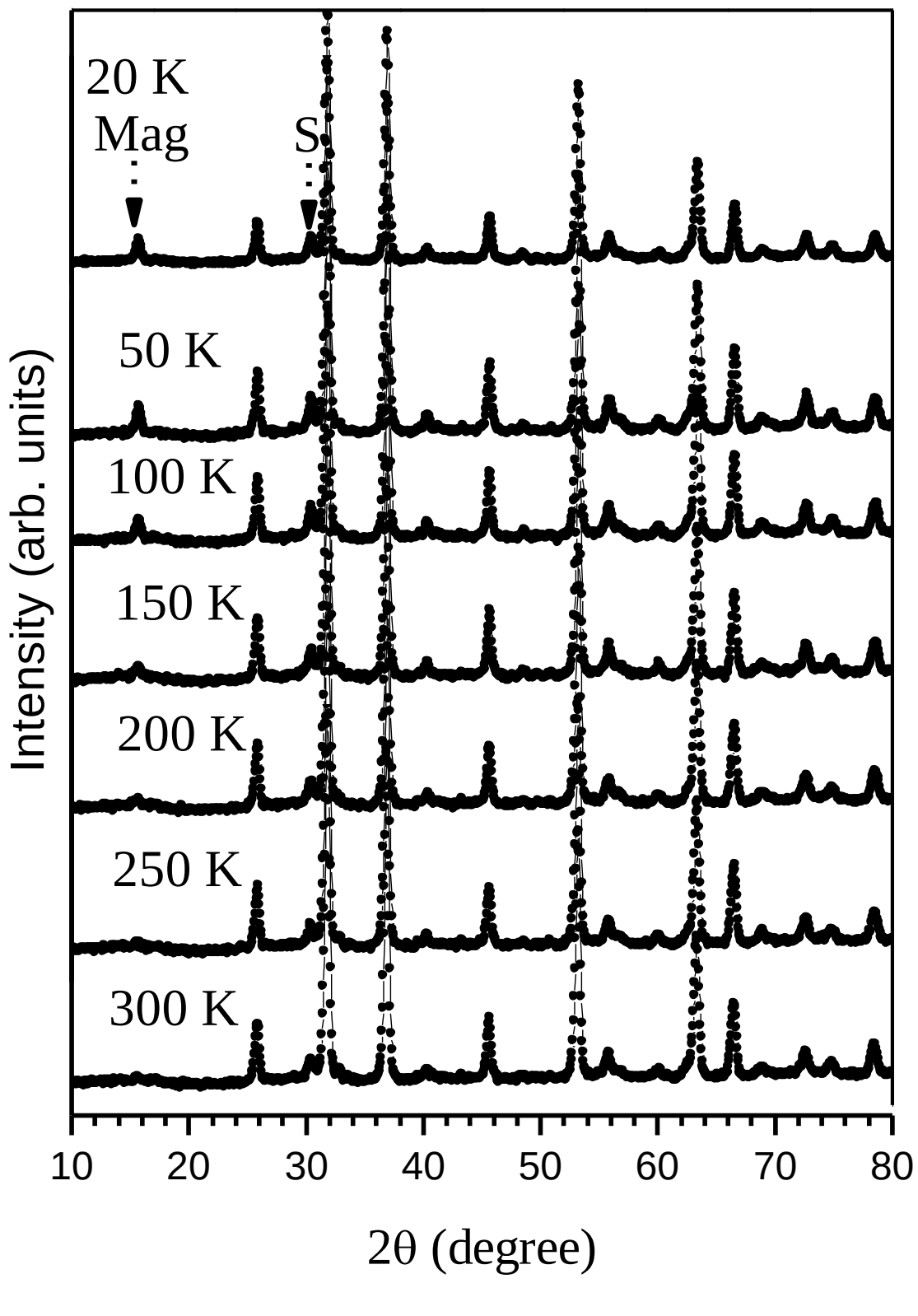

**Fig. 2** Evolution of the neutron diffraction patterns for BF-LF-PT with temperature for heating cycle. The superlattice(S) and magnetic (mag) peaks are marked with arrows.

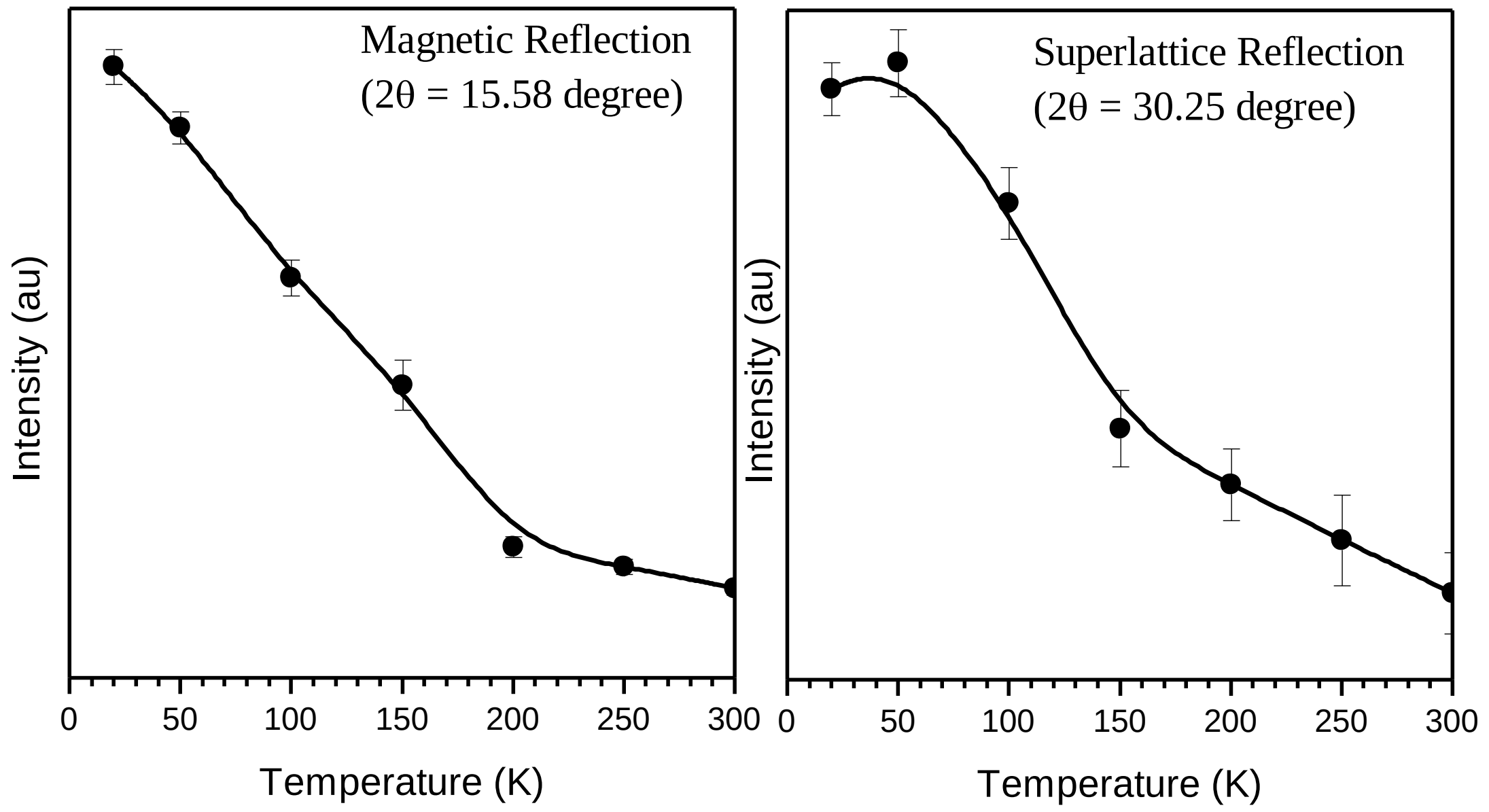

**Fig. 3** The variation of the integrated intensity of the superlattice (S) and magnetic (Mag) reflections appeared around 2θ =30.3 and 15.6 degrees with temperature.

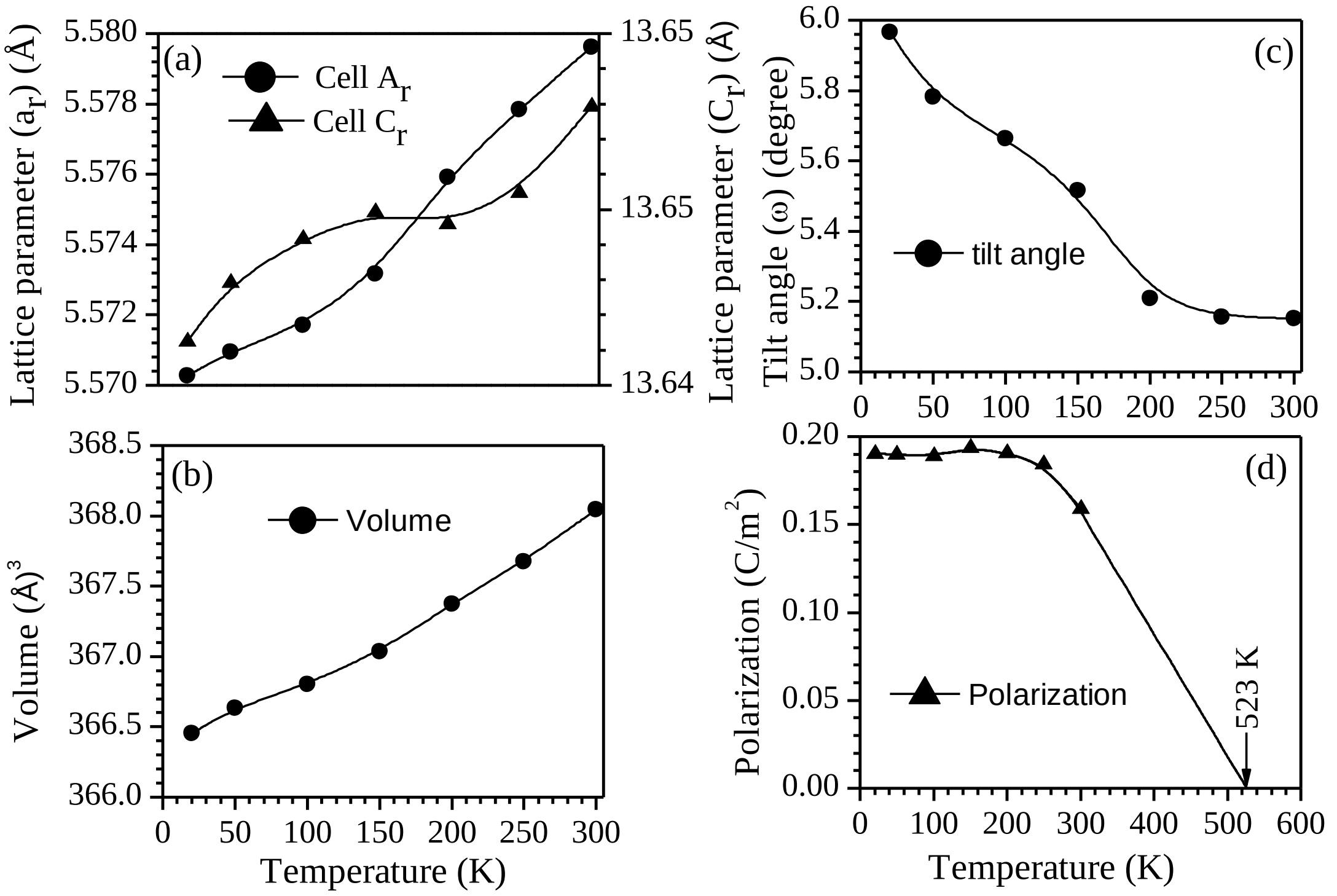

**Fig. 4** Evolution of (a) the calculated equivalent lattice parameters, (b) volume, (c) tilt angle and (d)calculated polarization obtained, after Rietveld refinements with temperature.

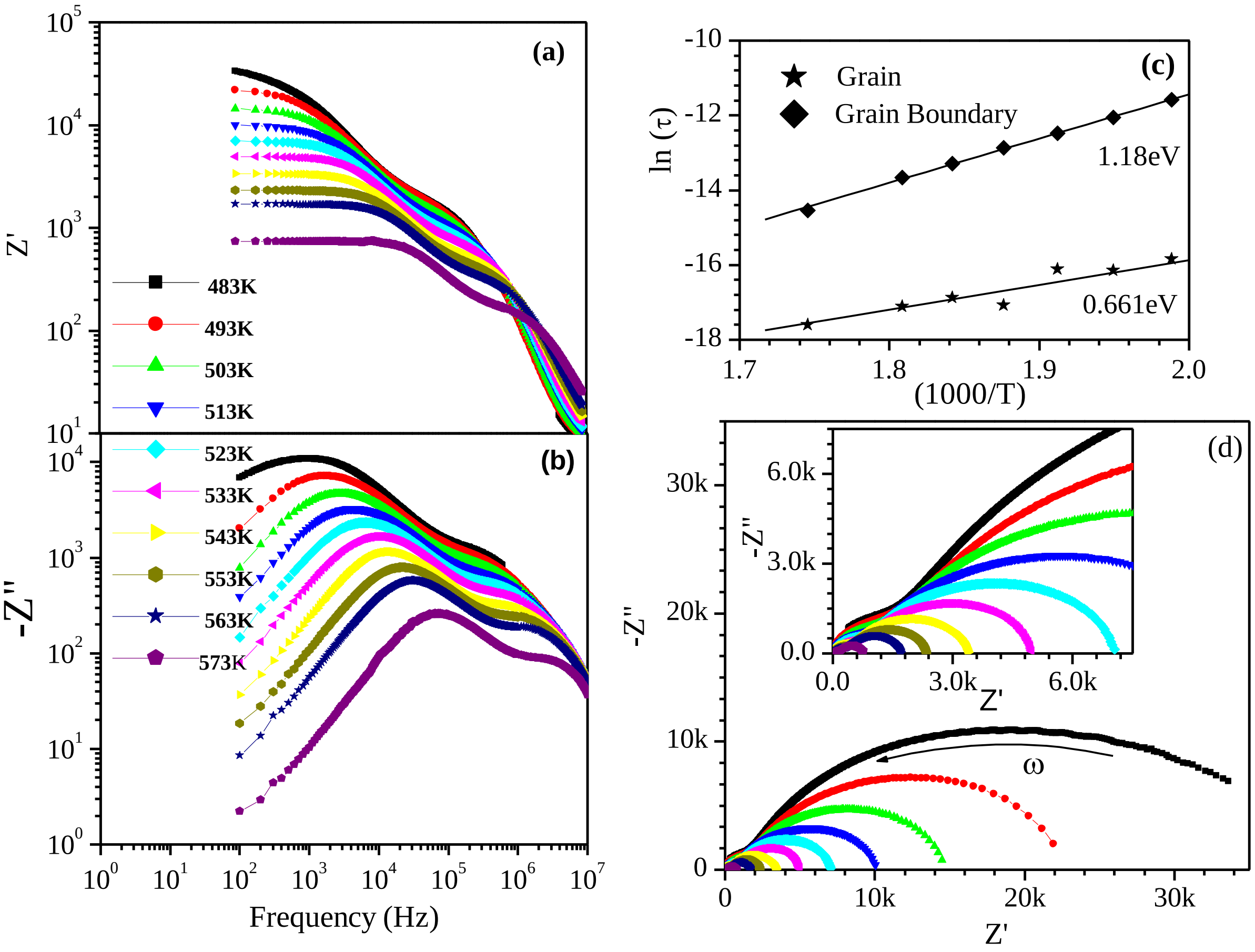

**Fig. 5** (Color online)The variations of the (a) real ($Z'$) and (b) imaginary parts ($Z''$) of impedance as a function of frequency at different temperatures (473 K≤ T ≤ 573 K). The activation energy associated with both (grain and grain boundary) relaxation processes is determined by using the fitting as shown in (c). Complex impedance plane plots $Z''$ vs. $Z'$ for BF-LF-PT are shown in (d). The inset shows that it is having two semi circle arcs.

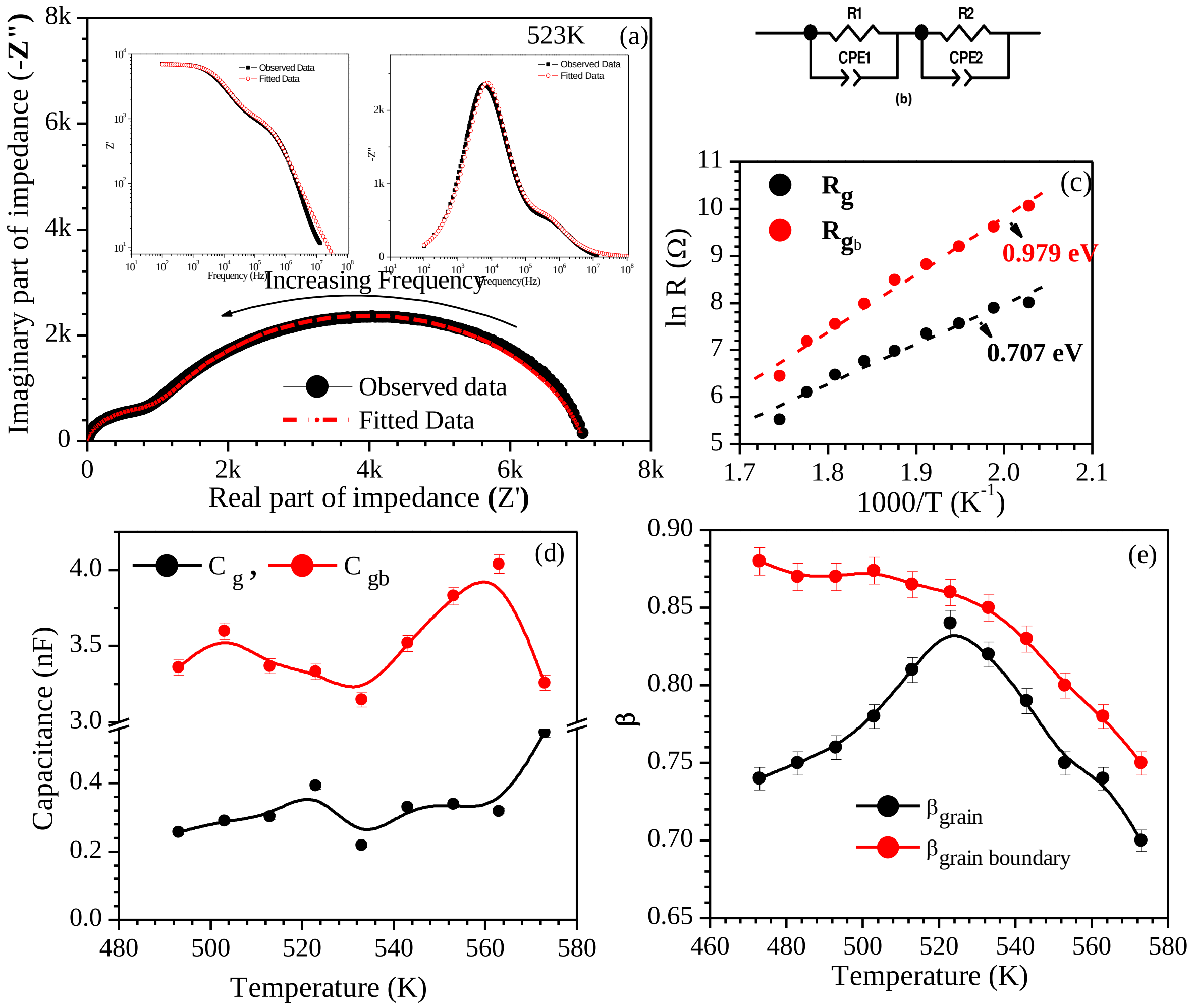

**Fig. 6** (Color online) (a) Complex impedance plane plots $Z''$ vs. $Z'$ fitted with R-CPE model for both grain and grain boundaries. (b) An equivalent circuit used to represent the electrical properties of grain and grain boundary effects. Variation of the fitting parameters resistance (R), capacitance (C) and (β) parameter used to calculate the deviation from the ideal Debye- type relaxation for both grain $R_g$, $C_g$, $β_g$ and grain boundaries $R_{gb}$, $C_{gb}$, $β_{gb}$ with temperature are shown in (c), (d) and (e) respectively.

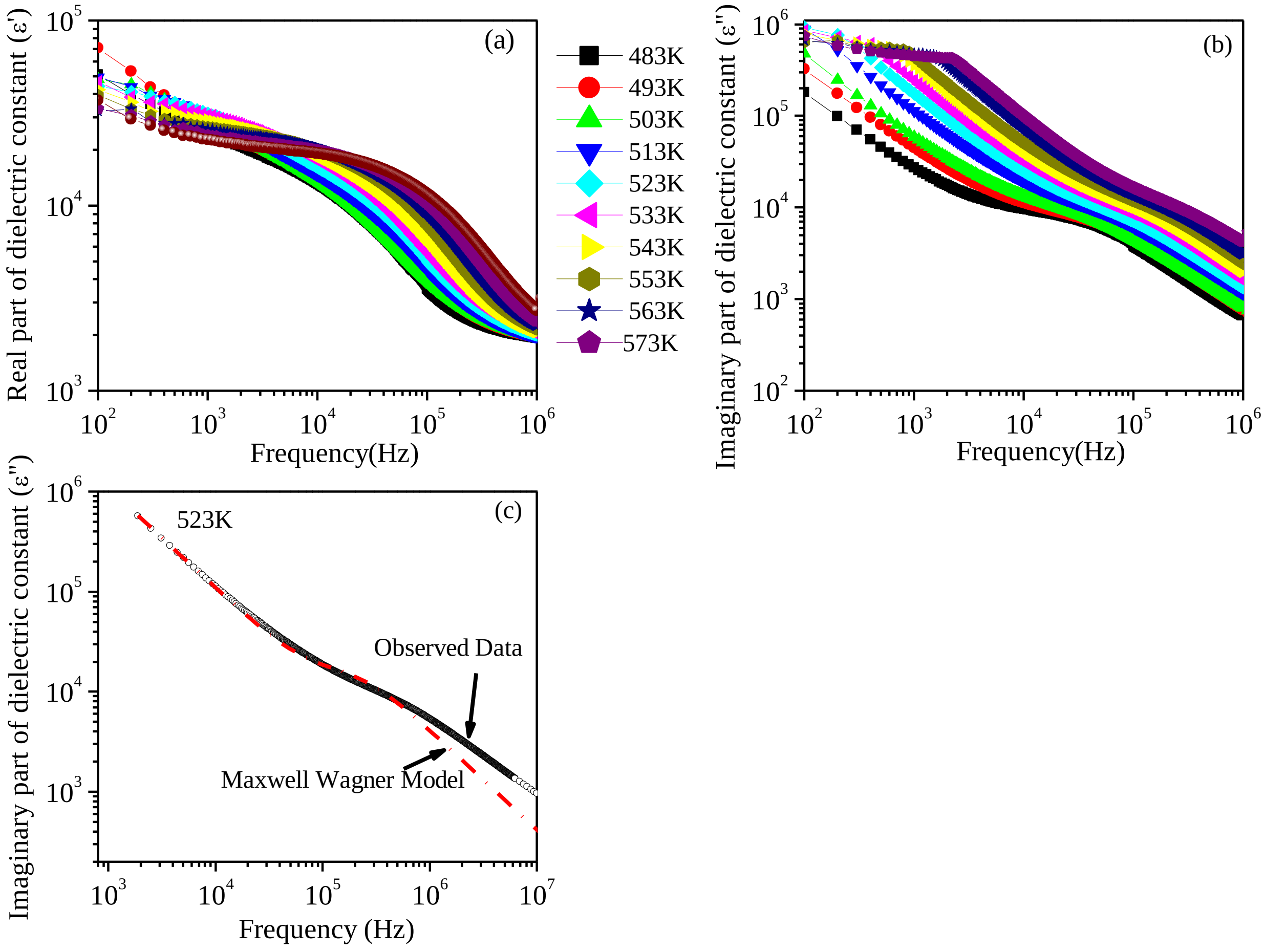

**Fig. 7** (Color online) Frequency dependence of (a) real (ε') and (b) imaginary (ε″) parts of dielectric constant at different temperatures (473 K≤ T ≤573 K). The fitting of Maxwell-Wagner model to imaginary (ε″) parts of dielectric constant vs frequency is shown in (c).

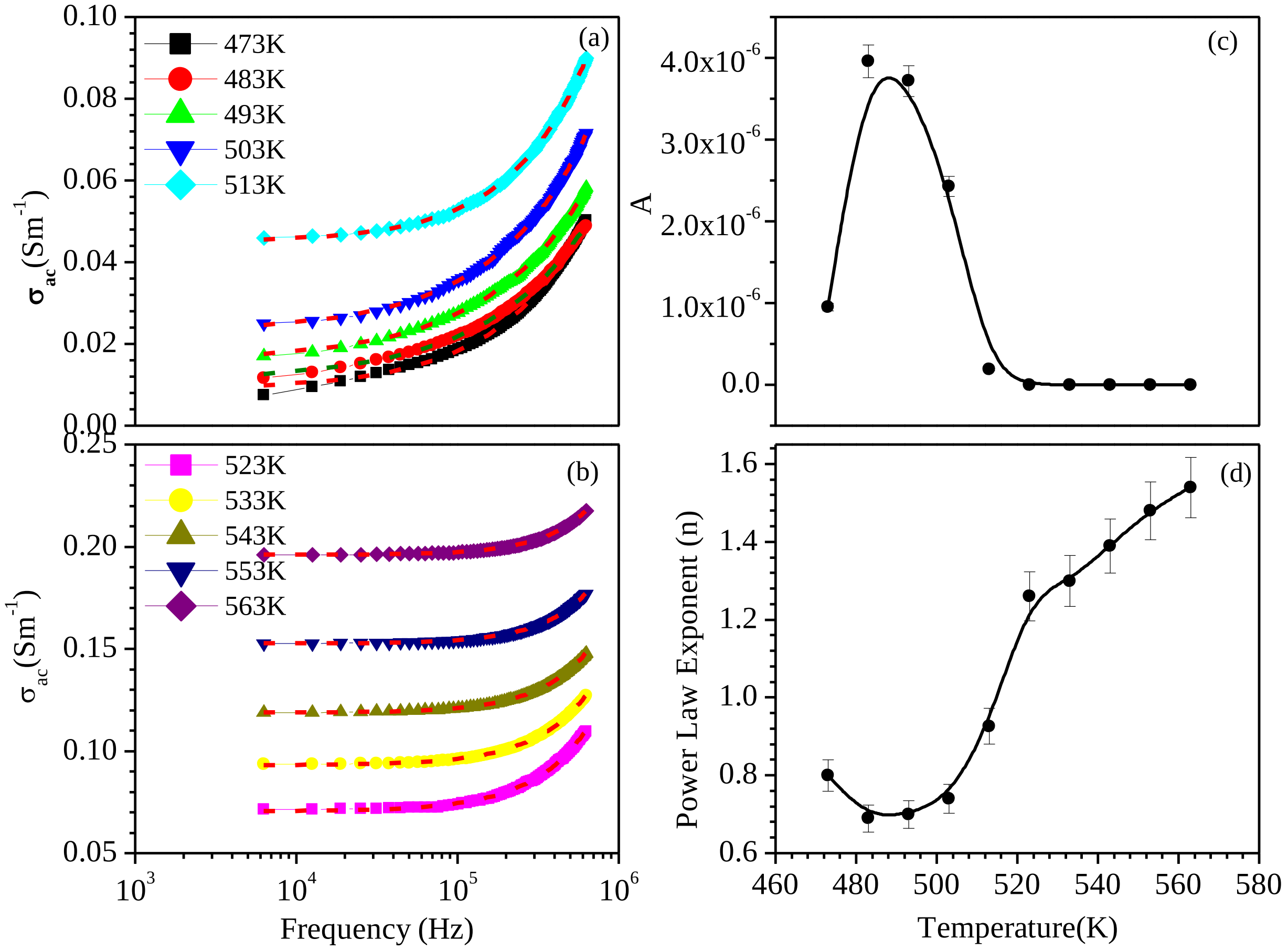

**Fig. 8** (Color online) Frequency dependence plot of the ac conductivity ( $\sigma_{dc}$) for temperatures ranging from 473 K to 573 K are shown in (a) and (b). The power law fits are shown as continuous line. The variation of power law pre factor A and exponent n as a function of temperature are shown in (c) and (d) respectively.

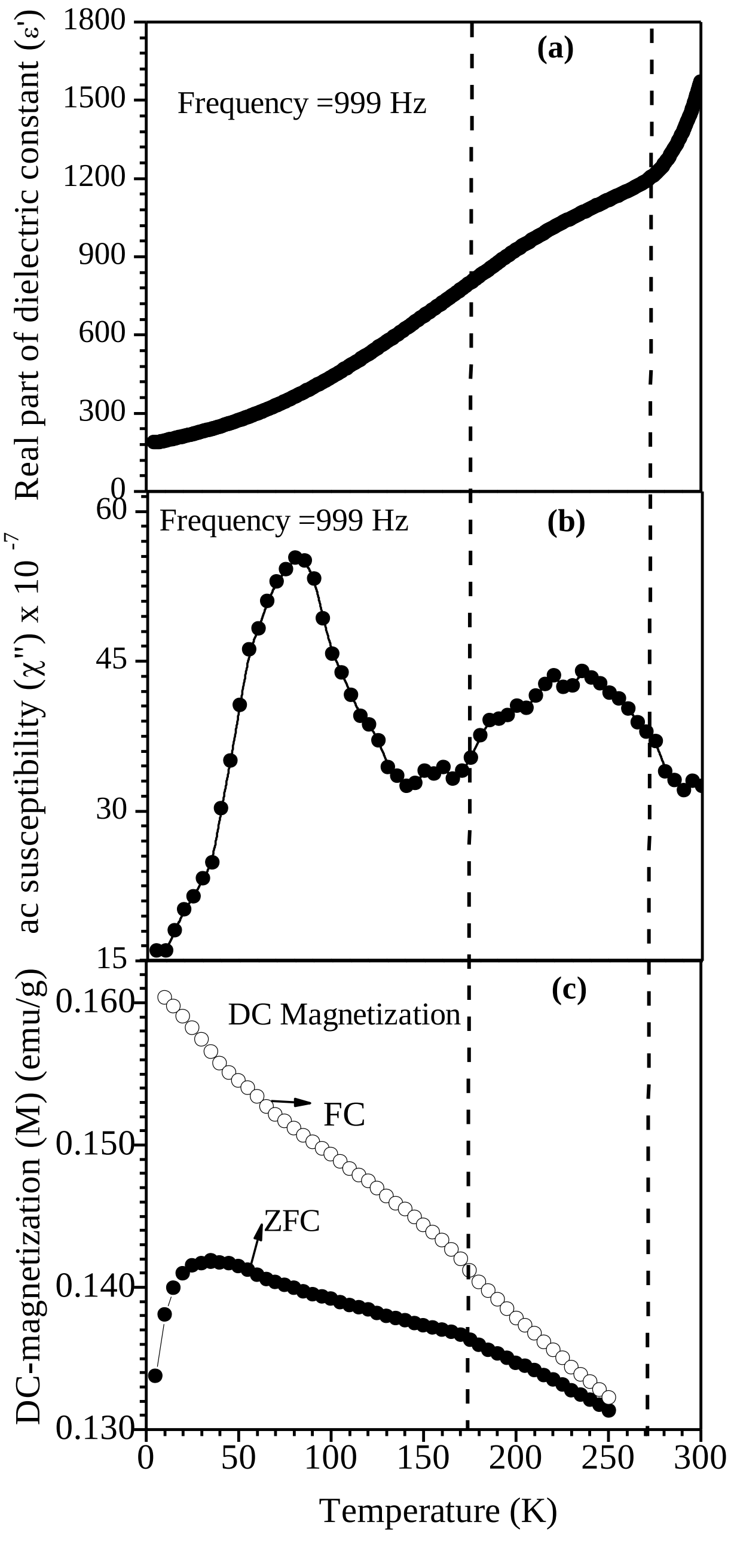

**Fig. 9** Temperature dependence of (a) dielectric constant (ε'), (b) ac susceptibility ( $\chi''$) and (c) dc magnetization (M) of BF-LF-PT.

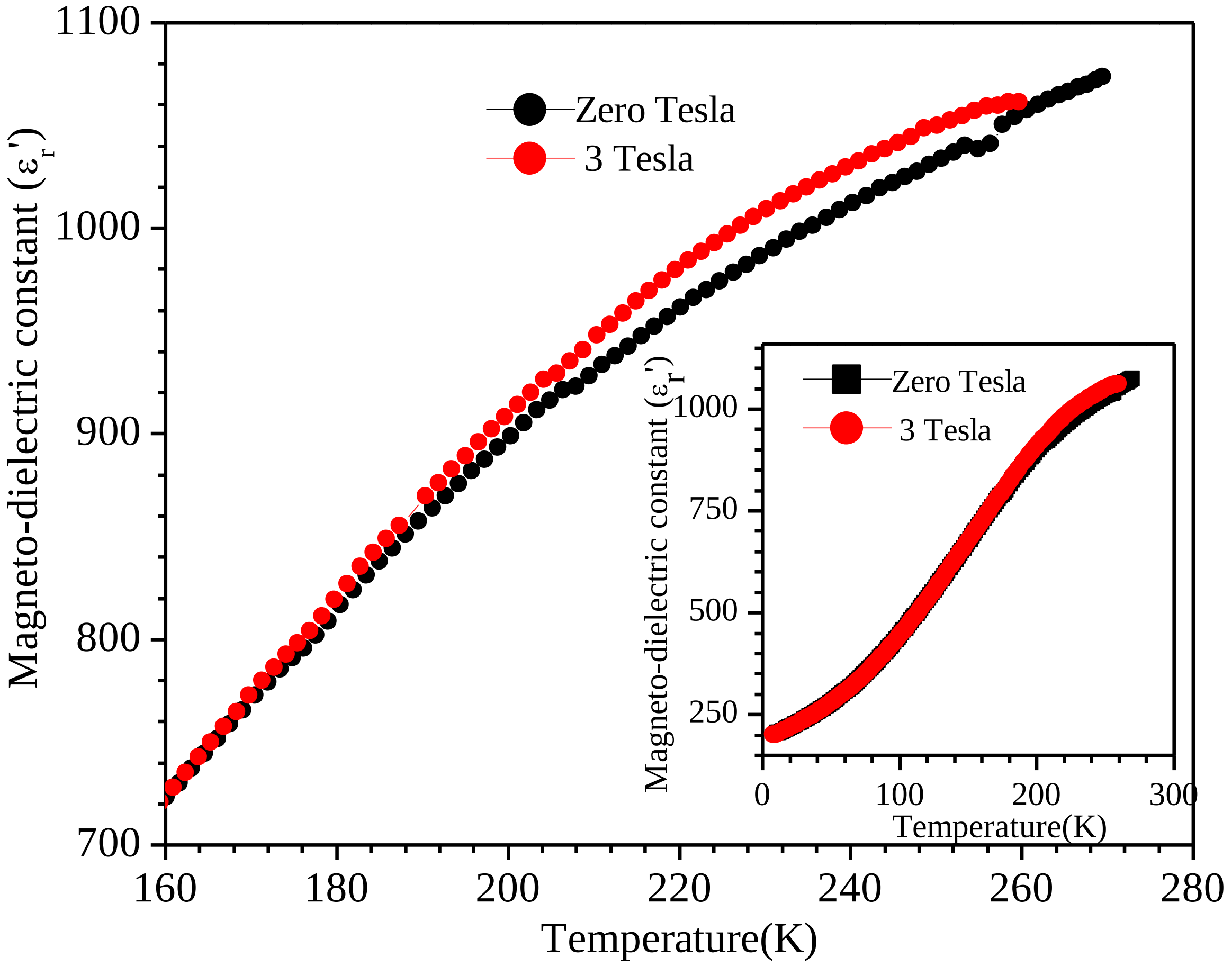

**Fig. 10** (Color online) Temperature dependence of Magneto-dielectric constant of BF-LF-PT.